\documentclass[journal=jctcce,layout=twocolumn,manuscript=article]{achemso}
\setkeys{acs}{articletitle = true}
\usepackage{amsmath}
\usepackage{bm}
\usepackage{dcolumn}
\newcolumntype{d}[1]{D{.}{.}{#1}}
\usepackage{graphicx}
\usepackage{multirow}
\usepackage{tabularx}
\usepackage{lineno}
\usepackage{txfonts}
\usepackage{booktabs}
\usepackage{subcaption}
\usepackage{gensymb}
\usepackage{siunitx}
\usepackage{adjustbox}
\usepackage[symbol]{footmisc}
\usepackage[usenames, dvipsnames]{color}
\usepackage{wrapfig} 
\usepackage{bm}
\usepackage{mathrsfs}
\usepackage{xurl}
\usepackage{hyperref}
\newcommand{\fig}[1]{Fig.~\ref{#1}}

\newcommand{\eq}[1]{Eq.~\ref{#1}}
\newcommand{\onlinecite}[1]{\hspace{-2 ex} \nocite{#1}\citenum{#1}}

\usepackage{stfloats}

\title{Stepping up enhanced rate calculations with EATR-flooding}

\let\oldmaketitle\maketitle
\let\maketitle\relax
\usepackage[fontsize=11pt]{fontsize}
\author{Nicodemo Mazzaferro}
\affiliation{Department of Chemistry, New York University, NY, 10003, USA}

\author{Willmor J. Pe\~na Ccoa}
\affiliation{Department of Chemistry, New York University, NY, 10003, USA}

\author{Pilar Cossio}
\email{pcossio@flatironinstitute.org}
\affiliation{Center for Computational Mathematics, Flatiron Institute, New York, 10010, USA}
\altaffiliation{Center for Computational Biology, Flatiron Institute, New York, 10010, USA}

\author{Glen M. Hocky}
\email{hockyg@nyu.edu}
\affiliation{Department of Chemistry, New York University, NY, 10003, USA}
\altaffiliation{Simons Center for Computational Physical Chemistry, New York University, NY, 10003, USA}

\begin{document}


\twocolumn[
\begin{@twocolumnfalse}
\oldmaketitle
\begin{abstract}
Several recent methods have shown that it is possible to compute rate constants of very slow biomolecular processes using simulations where a time-dependent bias is added along one or several collective variables (CVs). We previously reported the exponential average time-dependent rate (EATR) method, which can improve upon these approaches by accounting for how efficiently the external biasing potential modifies the observed rate using a learned CV-quality factor $\gamma$. This results in more accurate rate estimates using the same data when biasing a suboptimal coordinate. However, as formulated EATR depended on the biasing potential varying over time to properly determine the biasing efficiency, which limits the method's applicability to quasi-static biasing schemes such as ``flooding'' or on-the-fly probability enhanced sampling (OPES). Here, we present the EATR-flooding approach, which generalizes our method by replacing the need for a time dependent bias by instead varying (stepping up) the strength of the biasing potential across multiple sets of simulations. We implement this approach as an open-source Python library, and demonstrate that this approach is accurate without substantial loss of efficiency compared to standard EATR for a coarse-grained protein system, and also show good performance on a fully atomistic cavity-ligand model. Two additional appealing features of EATR-flooding are an internal check for over-biasing and the fact that only a single $\gamma$ parameter is predicted for a given choice of CVs, as compared to our earlier results where $\gamma$ empirically depended on biasing rate. Finally, we believe EATR-flooding applies not only to OPES simulations but more generally to CV biasing enhanced sampling approaches, making it broadly useful.
\end{abstract}
\end{@twocolumnfalse}]

\maketitle 

\section{Introduction}
The development of computational approaches to predict protein-ligand binding kinetics is an active area of research, due to the growing recognition of the importance of drug lifetime in controlling the efficacy and durability of pharmacological treatments \cite{ferruz2016binding,lotz2018unbiased,bernetti2019kinetics,nunes2020recent,ahn2021gaussian,smith2025toward}.
Classical molecular dynamics (MD) simulations provide a powerful approach to predict the behavior of biomolecules in solution as well as their interaction with ligands or other biomolecules \cite{karplus2002molecular,biophys_rev}.
Despite the name, MD simulations are more often used to estimate short timescale properties of a system by integrating the equations of motion to produce statistical samples of the system at fixed temperature or fixed temperature and pressure (distorting the microscopic kinetics), rather than long-timescale dynamical properties \cite{tuckerman2023statistical}.  
For many biophysical problems of interest, however, the dynamics of the most relevant configurational changes are dominated by transitions over free energy barriers between metastable states, meaning the kinetics can still be computed by statistical approaches.

Therefore, atomistic MD simulations can still provide a powerful technique for predicting the time scale or average rate constant for transitions between these states, as long as sufficient transitions can be observed within accessible computational time. 
Unfortunately, there is often a substantial difference in timescales between what can be accessed in MD simulations as compared to the time scales of the properties of interest.  Modern simulations often generate microseconds of trajectory data for relatively simple systems, whereas the processes of biomolecular interest occur on timescales ranging from milliseconds to seconds. 
 The transitions of interest are so-called rare events, in the sense that the system spends long periods in a metastable state before crossing into a new state in a relatively fast manner. 
Because of this undesired tendency to remain in a single region of state space for long periods of time, a number of enhanced sampling methods have been developed to quickly drive systems to escape free energy minima by adding an extra bias along certain collective coordinates or variables (CVs) describing the system \cite{henin2022enhanced,tuckerman2023statistical,kastner2011umbrella,metad,wtmetad,invernizzi2020rethinking,valsson2014variational}.
This extra bias effectively lowers the barrier to transitions, allowing one to reconstruct the underlying free energy landscape through inverting the effect of the bias on the frequency of observed configurations. 
The two such methods that we will focus on in this work are Metadynamics (MetaD) and On-the-fly probability enhanced sampling (OPES), which both work by constructing a bias on the fly which is tailored to push the system away from already explored regions of CV space \cite{metad,wtmetad,bussi2020using,invernizzi2020rethinking}. 

In both approaches, the consequence of adding this external bias is to significantly and intentionally reduce the long dwell times before the transitions. However, it is not always clear that rate constants of interest can be recovered under these biases in the same way as thermodynamic properties. 
Several methods have been developed to recover kinetics from CV biasing techniques,  relying on using a thermodynamic model for the kinetic process of interest, such as Kramers theory  \cite{voter1997hyperdynamics,imetad,imetadcdf,ray2022rare,ray2023kinetics,blumer2024short}.
In essence, doing so requires assuming that the bias is applied only to destabilize the starting state and does not directly affect the transition over the barrier, in which case transition times are sped up approximately  with the average of the exponential of the applied bias.
For example, in Infrequent Metadynamics (iMetaD) \cite{imetad}, this is achieved by only updating the biasing potential on a time scale longer than the barrier crossing time. In OPES-flooding, this is achieved by capping the highest level of bias applied to the system to fall below a previously estimated barrier level, $\Delta E$ \cite{ray2022rare}.
All of these methods also have an implicit assumption that the bias is applied to a CV that is a ``good'' reaction coordinate for the process of interest, since bias applied perpendicular to the ideal reaction coordinate should not be included in estimating such a speedup factor.
Despite these stringent requirements, iMetaD and OPES-flooding shine in the context of giving a computationally tractable way to estimate kinetics of very slow biomolecular processes (milliseconds to days) using only nanoseconds or microseconds of total simulation \cite{ray2023kinetics}.

In our work, we have concentrated on relaxing the requirement to choose a good CV, since in practice this is challenging to do \textit{a priori} - or even \textit{a posteriori}. We have pursued a strategy whereby we extract the rate constant from the data along with an additional parameter $\gamma$ that takes into account the efficiency of the chosen CV for producing desired transitions \cite{ktr,mazzaferro2024good}.
In our Exponential Average Time-dependent Rate (EATR) approach---described in detail in the next section---we estimate both the rate and $\gamma$ by averaging our estimate of the speed up factor across many MetaD simulations, and then fitting transition times either via maximum likelihood or by fitting the cumulative distribution of observed transition times. 

While EATR works well for simulations biased with MetaD, it failed when applied to simulations biased with OPES \cite{mazzaferro2024good}. 
This failure was due to the rapid convergence of OPES to a final biasing function in the starting basin, which prevents our fitting approach from disambiguating the unbiased rate from the estimate of CV quality, as explained below. 
To solve this, we propose a new method called EATR-flooding which estimates $\gamma$ using the biased rates from multiple sets of simulations, each with a different average amount of bias applied before transitions are observed; in OPES this is achieved by varying the value of $\Delta E$.
By ``stepping up'' the value of this parameter, which limits the average amount of bias that is added, we are able to probe the dependence of the observed rate on the amount of bias added, which can allow us to extract information about the quality of the biased CV(s) with minimal extra cost.
We first speculated about this approach in the supplementary information of our earlier study \cite{mazzaferro2024good} and showed how by varying the amount of bias, we could, with an uncontrolled approximation, extract an approximate rate and $\gamma$.
In this article, we put this method on more rigorous grounds.
We then test our new method on a coarse-grained model for protein folding as well as a fully atomistic model of a ligand binding problem, and show that it provides efficient and accurate  estimates for the unbiased rate, using a single $\gamma$ parameter to account for the quality of the biased CV.
We also show that the method generalizes beyond OPES, working for the case of a fixed external bias as originally envisioned in the conformational flooding or hyperdynamics methods \cite{grubmuller1995predicting, voter1997hyperdynamics}, and also for a time dependent bias applied with iMetaD.

The rest of this paper is organized as follows. In Sec.~\ref{sec:background}, we summarize the previously studied methods relevant to this article. In Sec.~\ref{sec:EATR-flooding}, we explain the necessity for and the theory behind the EATR-flooding method. In Sec.~\ref{sec:results}, we present the results of applying the EATR-flooding method to a coarse-grained protein and model cavity-ligand system. In Sec.~\ref{sec:conclusions}, we present our conclusions from the results. Finally, in Sec.~\ref{sec:methods}, we provide details of our computational methods.

\section{Background}
\label{sec:background}

\subsection{On-the-fly Probability Enhanced Sampling (OPES)}

OPES is an enhanced sampling technique that adds an additional potential energy term $V(\xi)$ called a \textit{biasing potential}, or simply the \textit{bias},  which depends on a CV $\xi$.  The biasing potential at time $t$ is given by
\begin{equation}
    V_t(\xi) = \left(1-1/B\right)\frac{1}{\beta}\log\left(\frac{P_t(\xi)}{Z_t}+\epsilon\right)\,,
    \label{eq:opesbias}
\end{equation}
where $B$ is the ``bias factor," $P_t(\xi)$ is the estimate at $t$ of the probability density, $Z_t$ is a normalization constant for $P_t(\xi)$, and $\epsilon=e^{-\beta\Delta E/(1-1/B)}$ is a regularization term included to hinder the bias increasing past a preset amount of energy $\Delta E$, termed \texttt{BARRIER} in the PLUMED open source sampling library \cite{tribello2014plumed,bonomi2019promoting}, as this is useful to keep the bias energy below an energy barrier for kinetics calculations.

In OPES, it is possible to enforce that bias is not applied when $\xi$ is within a pre-defined area, set by defining an ``excluded region." This can be useful for kinetics calculations, since this makes it possible to prevent bias from being added to CV values believed to capture the transition state \cite{ray2022rare}.
We do not necessarily need to use the excluded region for our calculations in this work, and indeed we did not use it for our more challenging atomistic system below.

OPES is particularly efficient in generating an estimate of the probability density due to the fact that $V(\xi)$ converges rapidly. For the OPES-flooding scenario where biasing is limited to the starting basin, after a short period at the beginning of the sampling, the bias remains effectively constant in time.

\subsection{Infrequent Metadynamics (iMetaD) and OPES-flooding (OPESf)}

iMetaD is a rate estimation technique inspired by hyperdynamics and conformational flooding \cite{imetad,voter1997method,grubmuller1995predicting}. For this method, a biasing potential is constructed as a function of a CV $\xi$ using MetaD, where a Gaussian of a specified height and width is added to the potential at a specified interval during an MD simulation. Several of these MetaD simulations are run until the transition occurs, and the transition time is rescaled according to
\begin{equation}
    \tau_i=\int_0^{t_i}e^{\beta V(\xi(t'),t')}\,dt'\,,
    \label{eq:imetad_rate}
\end{equation}
where $\tau_i$ and $t_i$ are the rescaled and observed transition times from simulation $i$, respectively \cite{imetad}. This is often expressed as $\tau_i=\alpha_it_i$, where
\begin{equation}
    \alpha_i=\frac{1}{t_i}\int_0^{t_i}e^{\beta V(\xi(t'),t')}\,dt'\,.
    \label{eq:imetad_acc}
\end{equation}
$\alpha$ is called the acceleration factor.

With the rescaled times, it is possible to estimate the rate, either by computing the inverse of the mean first passage time or by fitting the Poisson process cumulative distribution function (CDF) $h(\tau)=1-e^{-k_0\tau}$, using the rescaled times $\tau_i$ \cite{imetadcdf}.

OPESf follows the same principle as iMetaD, but applies it to simulations biased using OPES instead of MetaD\cite{ray2022rare}. This approach could offer advantages due to the ability to have finer control over the amount and location of added bias, as described in the previous section. 

\subsection{Exponential Average Time-dependent Rate (EATR)}
Both iMetaD and OPESf assume an ideal reaction coordinate such that all bias goes into accelerating transitions. For iMetaD, we previously reported on the KTR approach\cite{ktr} and EATR method\cite{mazzaferro2024good}, which allow us to compute more accurate rate constants from iMetaD trajectories by accounting for the fact that we are not biasing an ideal reaction coordinate.
Here, we focus on EATR which we demonstrated is a better ansatz, although KTR may still work equally well in real applications. 

In EATR, the instantaneous rate constant observed in the biased simulations is assumed to be related to the unbiased rate constant by the relationship
\begin{equation}
    k(t)=k_0 \overline{e^{\beta\gamma V(\xi(t),t)}}\,,
    \label{eq:eatr_rate}
\end{equation}
where $k(t)$ is the biased rate at time $t$, $k_0$ is the unbiased rate, $\beta=1/k_BT$ is the inverse thermodynamic temperature, $V(t)$ is the biasing potential evaluated at time $t$ for the value of the collective variable $\xi(t)$, and $\gamma\in[0,1]$ is the biasing efficiency measure from KTR. The overline represents an arithmetic mean over all simulations which did not transition before time $t$.

Because the rate constant is time-dependent, we model the transition as a nonhomogeneous quasi-static Poisson process with the survival probability
\begin{equation}
    S(t)=e^{-\int_0^tk(t')\,dt'}=e^{-k_0 \int_0^t\overline{e^{\beta\gamma V(\xi(t'),t')}}\,dt'}\,.
    \label{eq:eatr_surv}
\end{equation}

\begin{figure*}[t!]
    \centering
    \includegraphics[width=\textwidth]{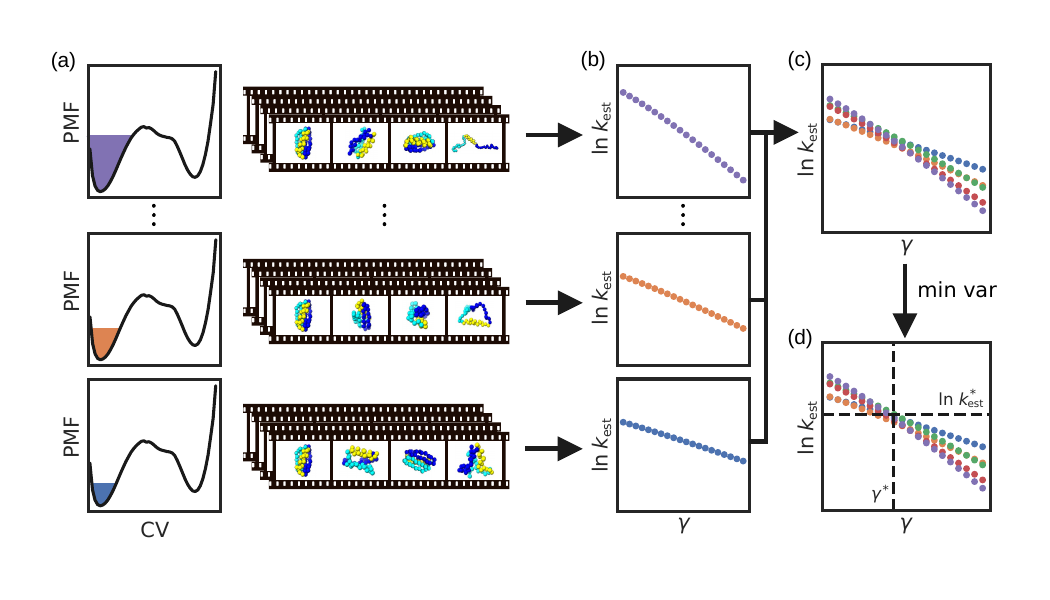}
    \caption{A schematic diagram for the EATR-flooding method. (a) Multiple sets of simulations are run until a transition occurs, where each set is run with a different average amount of applied bias, controlled by $\Delta E$ in OPES. Sets are color-coded for clarity. (b) The estimated log-rate for the transition  is computed as a function of $\gamma$,  $\ln k_\mathrm{est}  = \ln k_\mathrm{obs} - \ln\left<e^{\beta\gamma V(\xi)}\right>$. (c) The rate estimates are combined for all sets, and (d) the log-rate and $\gamma$ which minimize the variance over the sets are taken to be the EATR-flooding estimates.}
    \label{fig:eatrf_diagram}
\end{figure*}

If we perform $N$ biased simulations, where the first passage time of simulation $i$ is $t_i$, we can construct an empirical CDF $\hat{h}(t)=\frac{1}{N}\sum_{i=1}^N \Theta(t-t_i)$ where $\Theta(t)$ is the Heaviside step function. The theoretical CDF, $h(t)=1-S(t)$, can be fit to $\hat{h}(t)$ with the parameters $k_0$ and $\gamma$. The best-fit values of these parameters give us an estimate for the unbiased rate and the biasing efficiency, respectively.

If the biasing potential is constant in time, the simulation average in \eq{eq:eatr_rate} is approximately the ensemble average:
\begin{equation}
    \overline{e^{\beta\gamma V(\xi(t))}}\approx\left<e^{\beta\gamma V(\xi)}\right>\equiv\alpha_\gamma\,,
    \label{eq:const_bias}
\end{equation}
which is constant. In this case, the transition becomes Poissonian:
\begin{equation}
    S(t)=e^{-k_0 \int_0^t\left<e^{\beta\gamma V(\xi)}\right>\,dt'}=e^{-k_0 \alpha_\gamma t}\,,
    \label{eq:const_bias_surv}
\end{equation}
and there is not a unique combination of $k_0$ and $\gamma$ consistent with the data, since any decrease to $\alpha_\gamma$ can be compensated for by an increase in $k_0$. 
As such, EATR as previously formulated struggles to properly estimate $k_0$ and $\gamma$ when employing a static (or quasi-static) biasing potential, such as that used in OPES.

\section{EATR-flooding (EATRf)}
\label{sec:EATR-flooding}
In order to obtain estimates for $k_0$ and $\gamma$ using OPES, we need a formalism which does not assume that the biasing potential is time-dependent. We can do this by first applying \eq{eq:const_bias} to \eq{eq:eatr_rate}, taking the logarithm on both sides, and solving for the logarithm of the estimated rate constant, yielding
\begin{equation}
    \ln k_\mathrm{est}(\gamma) \equiv \ln k_\mathrm{obs} - \ln\left<e^{\beta\gamma V(\xi)}\right>\,,
    \label{eq:opes_flooding_k0}
\end{equation}
where $k_\mathrm{est}$ is the estimate for $k_0$, and $k_\mathrm{obs}$ is the constant value of the biased rate $k(t)$. Note that this rate estimate depends on $\gamma$. 

Given that there is only one true value of $\ln k_0$, we hypothesize that there exists a corresponding unique value for $\gamma^*$ which produces only one value of $\ln k_0^*$.  Therefore, we propose to determine $\gamma^*$ and $k_0^*$ by comparing the results of multiple \textit{sets} of simulations, each with a different amount of bias, as shown in \fig{fig:eatrf_diagram}a, and computing the effect of this different amount of bias on the observed rate constant.
The unbiased rate for each set is estimated as a function of $\gamma$ using \eq{eq:opes_flooding_k0}, then the unbiased rate and $\gamma$ estimates for the full collection of simulation sets are obtained by finding the value of $\gamma$ where all sets of simulations have approximately the same unbiased rate estimate, or equivalently, 
\begin{equation}
    \gamma^* = \mathrm{argmin}_\gamma \mathrm{Var}(\ln k_\mathrm{est}(\gamma))~,
\end{equation}
when the variance of the estimates across the sets is minimized. 
The best estimated value of the true rate is then $k^*_0=\overline{k_\mathrm{est}(\gamma^*)}$, where the average is done over the simulation sets.
This is illustrated in \fig{fig:eatrf_diagram}b-d.

We note that this approach does not require that the bias potential is static or quasi-static, but only that it is possible to converge an acceleration factor $\alpha_\gamma$ for each set of biasing parameters. 
Our EATRf method is therefore based on the same ansatz as the original EATR approach but is more general. 

\section{Results and Discussion}
\label{sec:results}
\subsection{Coarse-grained model for protein folding}
To evaluate our hypothesis in the preceding section, we first test our approach on a system sufficiently complicated that we can bias a wide range of CVs to obtain kinetic information, but sufficiently simple that we can obtain an unbiased rate constant estimate. 
We therefore adopted the same system used in our prior development of EATR, a G\={o}-like model of the B1 domain of protein G, which had previously been established as a useful system for studying the quality of CVs for describing protein folding.\cite{protg1,protg2,mazzaferro2024good}. As described in the Methods, we ran MD simulations where an OPES bias was applied to one of the following CVs: the fraction of native contacts $Q$, the end-to-end distance $R_\mathrm{ee}$, and the radius of gyration $R_\mathrm{g}$. We also performed simulations in which we biased $R_\mathrm{ee}$ and $R_\mathrm{g}$ simultaneously.

\begin{figure}[t]
    \centering
    \includegraphics[width=0.9
    \linewidth]{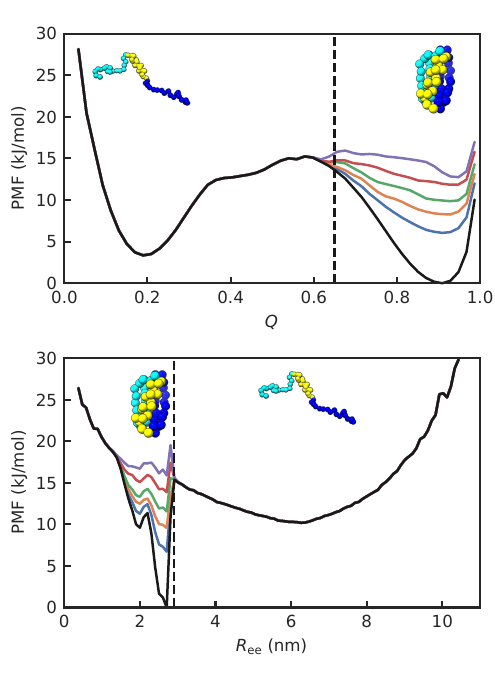}
    \caption{The potential of mean force (PMF) along the fraction of native contacts CV $Q$ (top) and the end-to-end distance CV $R_\mathrm{ee}$ (bottom). The colored lines indicate the PMF with the average final biasing potential from OPES with various values of the $\Delta E$ parameter. From bottom to top,  $\Delta E=$ 5, 7, 9, 11, and 13 kJ/mol. The vertical dashed lines represent the boundary of the excluded region.}
    \label{fig:opes_bias_heights}
\end{figure}

\begin{figure*}[t]
    \centering
    \includegraphics[width=\textwidth]{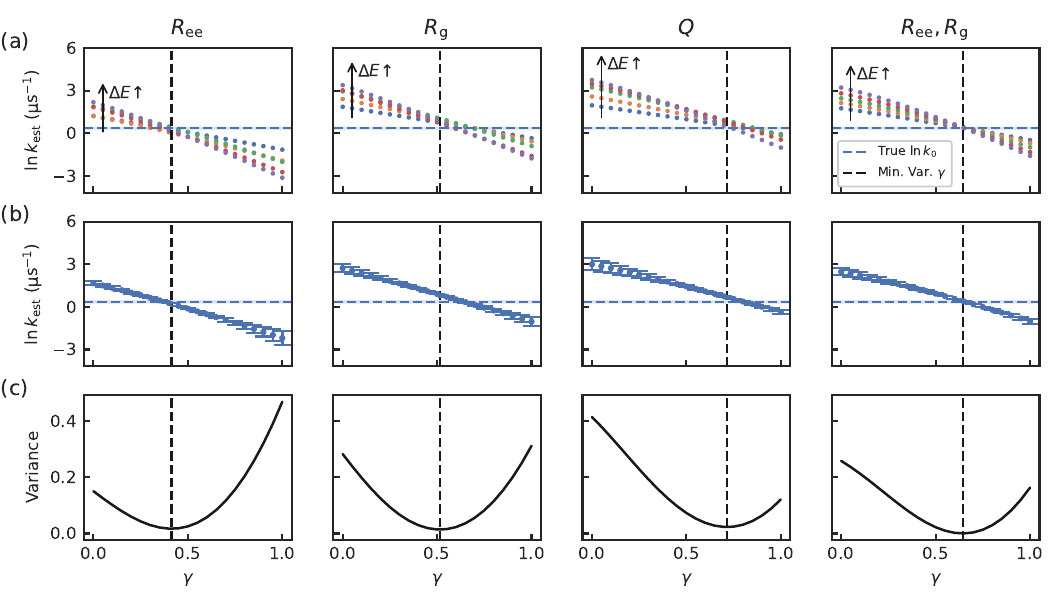}
    \caption{The results of EATR-flooding on the protein G model. (a) The values of $\ln k_0$ predicted by \eq{eq:opes_flooding_k0} for each set of OPES simulations with each $\Delta E$ parameter for various values of $\gamma$. (b) The value of $\ln k_0$ averaged over the sets of simulations for each potential value of $\gamma$. The variance of $\ln k_0$ across the sets is represented by error bars. (c) The variance in the predicted value of $\ln k_0$ across the sets of simulations. The horizontal dashed lines represent the true value of $\ln k_0$, the uncertainty in which is too small to represent in this plot. The vertical dashed lines represent the minimum variance value of $\gamma$, which is the value which causes all sets to give the same value of $\ln k_0$.}
    \label{fig:opes_gamma_scan}
\end{figure*}

For each choice of CV, we ran five sets of simulations where we varied the maximum level of OPES bias, $\Delta E$. This resulted in a different biasing potential in each set. 
As an example of how this looks when projected along two of the CVs, in Fig.~\ref{fig:opes_bias_heights} we plot the average final bias obtained from OPES on top of the underlying PMF computed in our previous work \cite{mazzaferro2024good}. As can be seen, the amount of bias/forcing increases in concert with the $\Delta E$ parameter, and depends on the choice of CV. 
In Fig.~\ref{fig:opes_biasmeasure}, we show that the effective average bias quickly plateaus to a static value, showing why we are not able to apply our original EATR method.

We therefore proceed to apply \eq{eq:opes_flooding_k0} to these data while varying $\gamma$ over the range 0--1, which gives us the results in \fig{fig:opes_gamma_scan}a. In this Figure, the horizontal dashed line shows the log of the unbiased rate computed previously \cite{mazzaferro2024good}.
With these data, we can directly see that the curves for each value of $\Delta E$ cross near the true rate. 
The crossover behavior can be explained as follows:
for $\gamma=0$, simulations with high $\Delta E$ are on top, which is simply a reflection of having faster observed transition times when the system is biased more strongly;
in contrast, $\gamma=1$ is equivalent to iMetaD which over-predicts transition times for non-ideal CVs \cite{mazzaferro2024good,mazzaferro2025using} and consequently larger over-predictions (lower rate constants) arise for higher $\Delta E$. 

In \fig{fig:opes_gamma_scan}b,c, we show the average and variance across data sets obtained from the values computed via \eq{eq:opes_flooding_k0}.
With a vertical dashed line, we indicate the $\gamma$ which has the lowest variance, as computed using an optimization algorithm as described in the methods. 
Visually, we can observe that the minimum variance occurs close to when the estimated log rate value intersects that of the unbiased rate estimate. 

Inspecting the minimum-variance $\gamma$ values, we find that they are similar to those computed using our original EATR approach using data from iMetaD \cite{mazzaferro2024good}.
As expected, $Q$ has the highest value of $\gamma$ while $R_\mathrm{ee}$ has the lowest. Simultaneously biasing the two lower quality CVs, which individually have $\gamma$ values of 0.42 and 0.54, results in a combined $\gamma$ value of 0.68, showing the synergistic value of multidimensional biasing. 
Of note, the $\gamma$ value for $Q$ is somewhat less for these OPES simulations than we observed with iMetaD, where $\gamma$ could approach 1 for slow biasing along $Q$, which may reflect an interesting difference between how OPES and iMetaD build up the bias histogram, which is worthy of further investigation in the future.

In addition to this $\gamma$ analysis, we can also see the quality of the CVs visually by plotting the log of the observed rate versus the log acceleration factor, $\ln \langle e^{\beta V}\rangle$, as shown in Fig.~\ref{fig:opes_slope_protg}. 
Since the OPESf acceleration factor is accurate for good CVs such as $Q$, we expect a linear relationship with near unit slope between these two terms.
As discussed in Ref.~\citenum{mazzaferro2024good}, lower slopes indicate a bad or inefficient CV. 
This is discussed more in the next section.

\begin{table}[h]
   \begin{center}
    \caption{\label{tab:protgresults} The rates and $\gamma$ values obtained for the protein G model by applying EATR-flooding to different CVs, compared to the rates obtained from conventional OPES-flooding, combining data from all sets with the $\Delta E$ values 5, 7, 9, 11, and 13 kJ/mol. The `unbiased' entry is obtained from running 100 equilibrium MD simulations, from Ref.~\citenum{mazzaferro2024good}.}
    \begin{tabularx}{\columnwidth}{>{\raggedright\arraybackslash}X >{\centering\arraybackslash}X >{\centering\arraybackslash}X}
    \hline
    \multicolumn{3}{c}{\textbf{OPES-flooding}}   \\ 
     \hline
    CV & $\ln k_0^\mathrm{est}~\mathrm{(\mu s^{-1})}$  & \\
    \hline
    $R_\mathrm{ee}$  & $-1.92\pm0.07$ &   \\ 
    $R_\mathrm{g}$ & $-0.82\pm0.05$ &   \\ 
    $R_\mathrm{g}$, $R_\mathrm{ee}$ & $-0.63\pm0.05$ &   \\ 
    $Q$ & $0.02\pm0.05$ &   \\ 
    \hline\hline
    \multicolumn{3}{c}{\textbf{EATR-flooding (OPES)}} \\
    \hline
    CV & $\ln k_0^\mathrm{est}~\mathrm{(\mu s^{-1})}$ & $\gamma$ \\
    \hline
    $R_\mathrm{ee}$ & $0.23\pm0.23$ & $0.41\pm0.06$  \\
    $R_\mathrm{g}$ & $0.85\pm0.16$ & $0.52\pm0.04$  \\
    $R_\mathrm{g}$, $R_\mathrm{ee}$ & $0.42\pm0.17$ & $0.64\pm0.04$  \\
    $Q$ & $0.62\pm0.19$ & $0.74\pm0.06$  \\
    \hline\hline
    \multicolumn{3}{c}{\textbf{EATR-flooding (fixed bias)}}  \\
    \hline
    CV & $\ln k_0^\mathrm{est}~\mathrm{(\mu s^{-1})}$ & $\gamma$ \\
    \hline
    $R_\mathrm{ee}$ & $0.40\pm0.18$ & $0.31\pm0.07$  \\
    $Q$ & $0.53 \pm 0.16$ & $0.88\pm0.07$ \\
    \hline\hline
    \multicolumn{3}{c}{\textbf{EATR-flooding (iMetaD)}}  \\
    \hline
    CV & $\ln k_0^\mathrm{est}~\mathrm{(\mu s^{-1})}$ & $\gamma$ \\
    \hline
    $R_\mathrm{ee}$ & $0.43\pm0.06$ & $0.28\pm0.01$  \\
    $Q$ & $0.47\pm0.17$ & $0.67\pm0.06$ \\
    \hline\hline
    unbiased: & $0.36\pm0.07$ &  \\
    \end{tabularx}
    \end{center}
\end{table}

\begin{figure*}[h]
    \centering
    \includegraphics[width=\textwidth]{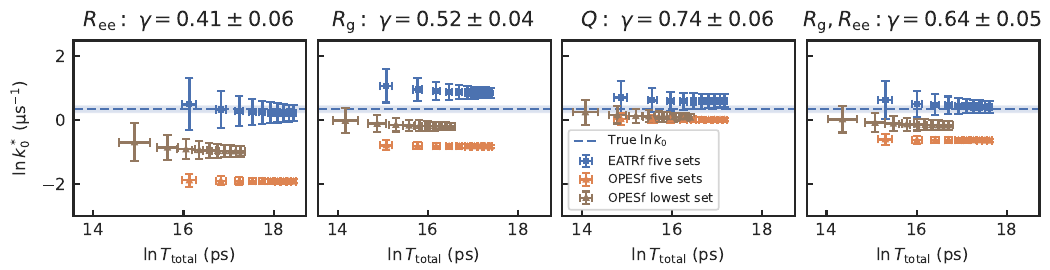}
    \caption{The predicted value of $\ln k_0$ from EATR-flooding (in blue) and OPES-flooding (in orange) as a function of the logarithm of the total simulation time with the lowest five values for $\Delta E$. Various numbers of simulations per set, from 5 (left side of each plot) to 100 (right side of each plot). The blue dashed line represents the rate obtained from unbiased simulations. Error bars indicate the standard error in bootstrap analysis.}
    \label{fig:opes_rate_vs_time}
\end{figure*}

Having demonstrated that EATR-flooding is an accurate approach, we then wanted to investigate whether the additional cost of running with multiple values of $\Delta E$ made the method prohibitively expensive compared to OPESf. 
To do so, we recomputed the predicted rates by EATRf using subsampling with smaller numbers of simulations per batch.
As shown in \fig{fig:opes_rate_vs_time}, the typical predicted rate constant is unchanged even when using only 5 simulations per batch, corresponding to 50x less simulation time than used for the full analysis. However, the typical error in doing so is likely to be larger, and so there is some benefit to obtaining more data. 
From this analysis, we directly observe that the OPESf estimator is systematically incorrect for poor CVs, and additional sampling does not alleviate the issue.
In contrast, the mean predicted value of $k_0$ from EATRf is generally close to the unbiased value for all CVs, always within a factor of 2, and seems to have near perfect accuracy for 2d biasing of $R_\mathrm{g}$ and $R_\mathrm{ee}$ (see also Tab.~\ref{tab:protgresults}). 

\begin{figure}[t]
    \centering
    \includegraphics[width=\linewidth]{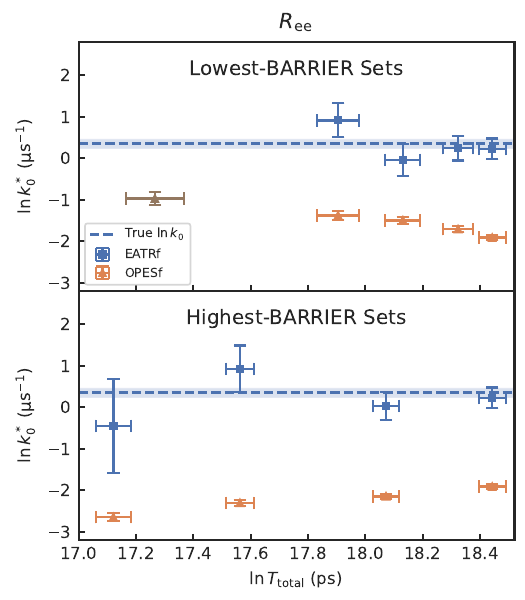}
    \caption{The predicted value of $\ln k_0$ from EATR-flooding and OPES-flooding on the $R_\mathrm{ee}$ CV with the lowest (top) and highest (bottom) values for $\Delta E$ with 100 simulations per set. The blue dashed line represents the rate obtained from unbiased simulations. Error bars indicate the standard error in bootstrap analysis.}
    \label{fig:opes_number_sets}
\end{figure}

As just discussed, \fig{fig:opes_rate_vs_time} shows that our approach of combining data from multiple sets of simulations biased by different amounts is successful and efficient. 
In that analysis, we always combined results from five different values of $\Delta E$, and showed that accurate predictions of the rate could be obtained even with only a small number of simulations per set. 
We also wanted to investigate what the minimum number of different $\Delta E$ values we could use is, and also whether it was required to have the lowest $\Delta E$ values (which require the most simulation time to observe transitions). 
\fig{fig:opes_number_sets} shows the result of this analysis, where for the worst CV, $R_\mathrm{ee}$, we systematically eliminate either 0, 1, 2, or 3 of our data sets, while maintaining a constant number of simulations within each data set. 
As was the case in \fig{fig:opes_rate_vs_time}, we see that the estimate from the OPES-flooding estimator is systematically wrong in all cases. 
Interestingly, as seen in the top panel, adding more data sets with higher $\Delta E$ makes the OPESf estimate worse; in the lower panel, we see the first point using only the highest two $\Delta E$ is particularly bad, but adding more data at lower barriers slightly improves the estimate. 
In stark contrast, using the same data the EATRf estimator is relatively accurate when using just the lowest 2 barriers, and becomes systematically better as more data is added. 
It is also 2 log units better than the OPESf estimate with even just the highest two $\Delta E$ sets included, albeit with a large variance. 

Given that EATRf was conceived of in the context of quasi-static biasing, we also performed simulations where we added a truly static Gaussian bias rather than using OPES. We performed otherwise identical simulations and analysis, adding a constant external potential to either $R_\mathrm{ee}$ or $Q$, described in  Sec.~\ref{sec:static_bias} in the SI.
\fig{fig:eatrf_static_bias} shows the EATRf analysis for these data, which shows equally or more accurate results as compared to the OPES data, despite the naive way in which the bias was constructed (Tab.~\ref{tab:lj}). 

Finally, to further demonstrate the generality of EATRf, we applied our workflow to data generated by iMetaD for Ref.~\citenum{mazzaferro2024good}, where the different forcing amounts are controlled by the frequency of hill deposition alone. As described in SI Sec.~\ref{sec:metad_bias} and seen in Fig.~\ref{fig:eatrf_metad_bias} and Tab.~\ref{tab:protgresults}, this approach is as accurate as our EATRf results using OPES. 

\subsection{Cavity-Ligand Model}
\begin{figure*}
    \centering
    \includegraphics[width=\textwidth]{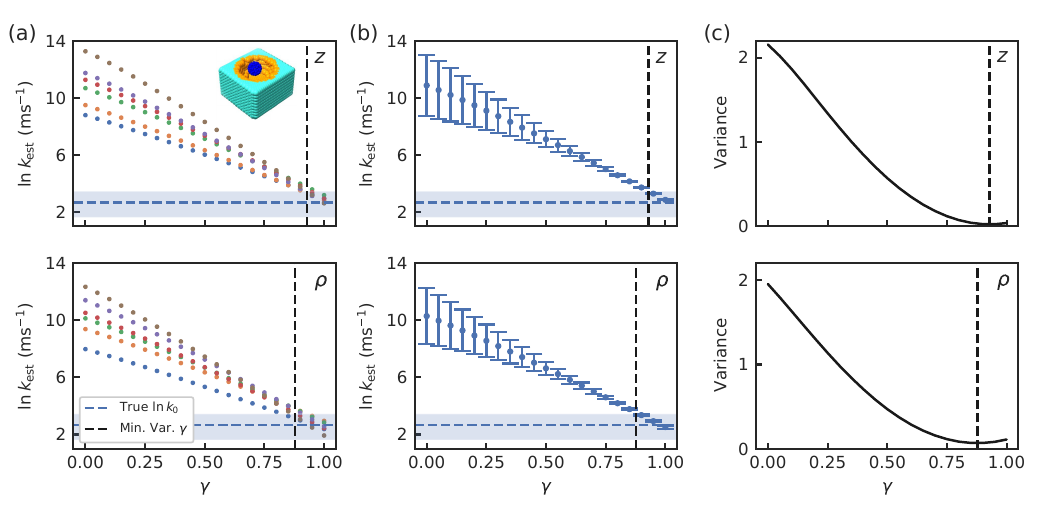}
    \caption{The results of EATR-flooding on the weakened cavity-ligand model. (a) The values of $\ln k_0$ predicted by \eq{eq:opes_flooding_k0} for each set of OPES simulations with each $\Delta E$ for various values of $\gamma$. (b) The value of $\ln k_0$ averaged over the sets for each value of $\gamma$. The variance of $\ln k_0$ across the sets is represented by error bars. (c) The variance in the predicted value of $\ln k_0$ across the sets of simulations. The horizontal dashed lines represent the true value of $\ln k_0$ with the 95\% credible interval. The vertical dashed lines represent the minimum variance value of $\gamma$.}
    \label{fig:opes_gamma_cavity_weak}
\end{figure*}
Although EATRf seems promising on a toy model, we want to ensure that it works for a larger and fully atomistic system. A persistent challenge is finding a challenging benchmark system that still admits a known ``ground truth'' and which also has multiple possible choices of CVs to bias. 
For this study, we adopted a simple model of receptor-ligand unbinding that was originally developed to study the effect of collective dewetting on ligand dissociation\cite{mondal2013hydrophobic}. 
This system consists of a buckyball in a hydrophobic cavity, and it is known that the typical unbinding reaction occurs with the buckyball sliding sideways in the pocket before escaping, which allows the pocket to become hydrated, rather than coming directly out which creates a temporary vacuum \cite{tiwary2015role}.
We have previously used this model to study the effect of mechanical pulling forces on the unbinding rates using iMetaD \cite{pena2022assessing}.
However, we do not know the ``true'' unbinding rate, and estimates from an alternative approach, Markov state modeling, gives an estimate orders of magnitude faster than that from iMetaD \cite{ahalawat2020solvent}.
For this study, we therefore decided to modify the parameters of this model to produce a series of cavity-ligand models with lower binding affinities, from which we can select a system that displays the same qualitative behavior but for which unbinding occurs fast enough to allow us to estimate an unbiased rate. 

As detailed in the SI, we reduced the $\epsilon$ parameters of the non-bonded interactions between the receptor and ligand atoms. Three different models with decreasing attractive interactions were evaluated. We computed the unbinding PMF using the MetaD procedure from our previous study\cite{pena2022assessing}, with results shown in  Fig.~\ref{fig:wcavitys}. 
As $\epsilon$ is lowered slightly (SI Tables \ref{tab:lj} and \ref{tab:ljnb}) we see a systematic decrease in the barrier to unbinding. 
The third model ($\epsilon_3$) tested has a barrier approximately 10 kcal/mol lower and we proceeded to assess whether we could approximate its unbinding rate constant with unbiased MD\cite{pena2022assessing}.
At the same time, as shown in Fig.~\ref{fig:wcavitys}a,b, the underlying free energy landscape has the same shape as in the original model, meaning we still expect unbinding to follow the same type of nontrivial unbinding pathway.

To estimate an unbiased rate constant for our model, we performed 360 total microseconds of unbiased MD launched from 48 different sampled starting points.
This produced only five transitions. 
Using these data we made a maximum likelihood estimate for the unbiased log-rate of 2.67 (with $k_0$ in $\mathrm{ms}^{-1}$), with a highest-density 95\% credible interval of 1.62 to 3.44 determined according to Sec.~\ref{sec:error}.

We then performed sets of 48 OPES simulations each for a series of $\Delta E$ values from 3 to 6 kcal/mol. 
We biased two CVs: the perpendicular ($z$) and orthogonal ($\rho=\sqrt{x^2 + y^2}$) distance from the ligand to the center of the cavity.
For larger and more complex systems like this, we first examined the change in the logarithm of the rate constant with the logarithm of the average acceleration factor, which is presented in \fig{fig:opes_slope_cavity}.
In practice, we have observed that when the system is not over-biased, this curve will be linear with a slope approximately equal to $\gamma$ (equivalent to taking $\gamma$ outside the average in Eq.~\ref{eq:opes_flooding_k0}). 
For very large $\Delta E$, we observe that the points start to curve down, and we take this as a heuristic of which points to exclude for our EATRf analysis, as these likely correspond to over-biasing regimes.

Taking 6 values of $\Delta E$ (3, 3.5, 4, 4.5, 5, and 6 kcal/mol) we performed the EATRf analysis with results shown in Fig.~\ref{fig:opes_gamma_cavity_weak}. When biasing the $z$-coordinate, we obtained $3.47\pm0.34$ as the estimate for $\ln k_0$ with $k_0$ in $\mathrm{ms}^{-1}$ and $\gamma=0.93\pm0.04$. When biasing the $\rho$-coordinate, we obtained $\ln k_0=3.50\pm0.36$ and $\gamma=0.88\pm0.04$. 
We find that $z$, the natural unbinding coordinate, has a high $\gamma$ and a predicted rate constant within the credible intervals of our unbiased estimate. 
For $\rho$, we get a slightly lower $\gamma$ as expected since it doesn't directly drive the unbinding. Using the traditional OPESf analysis from just our lowest barrier data produced $\ln k_0=2.89\pm0.06$ for $z$ and $\ln k_0=2.50\pm0.07$ for $\rho$.

Although we do not have an estimate of the unbiased rate for the original ``strong'' model, we did perform EATR-flooding analysis as shown in \fig{fig:opes_gamma_cavity} in Sec.~\ref{sec:si_cavity_results} of the SI. In this case, we find that $z$ alone is predicted to be a less ideal CV than in our weaker case ($\gamma=0.64$) and this is not substantially improved by biasing $z$ and $\rho$ simultaneously (also $\gamma=0.64$). The predicted unbinding times fall between those for MSM and previous iMetaD studies and are probably underestimates given our predicted difference in the barrier heights of the original and new models.

\begin{figure}[h!]
    \centering
    \includegraphics[width=\linewidth]{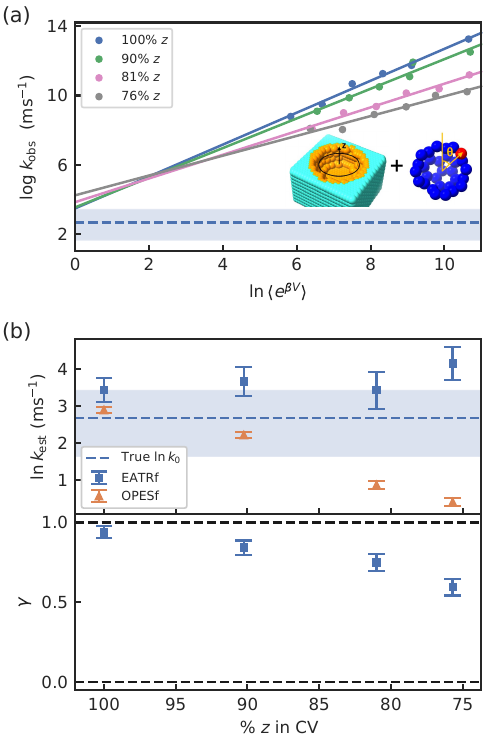}
    \caption{(a) The observed value of the log rate in the simulation sets while biasing a CV defined as linear combinations of $z$ and $\theta$, plotted against $\ln\left<e^{\beta V}\right>$, which acts as a measure of the amount of bias added. For worse CVs that have a larger contribution from $\theta$, adding the same amount of bias has less of an effect on the observed rate. Solid lines are linear fits to the data, with slopes from top to bottom of 0.94, 0.84, 0.75, and 0.59. (b) The predicted value of $\ln k_0$ (top) and $\gamma$ (bottom) from EATR-flooding and OPESf while biasing each of the CVs. The x-axis is the value of $a^2$ where $a$ is the coefficient of $z$ in the linear combination. Mean $\gamma$ values are 0.92, 0.86, 0.68, and 0.57, quite similar to the values of slope in (a), in accordance with the hypothesis in Ref.~\citenum{mazzaferro2024good}.  }
    \label{fig:cavity_lincombo}
\end{figure}

Returning to our weaker cavity model, because $z$ and $\rho$ were both found to be good CVs for biasing in this system, we decided to define a set of CVs to demonstrate that EATRf still works even if the quality of the CV is worse than that of $\rho$.
We first created a nuisance CV $\theta$ that represents the zenith angle of a chosen carbon atom (defined in PLUMED using a dihedral between that atom, the center of the buckyball, the center of the cavity, and an atom in the cavity that is aligned to the vertical axis), then we created four CVs, each of which is of the form $\xi=a\,z+b\,\theta$, where $a^2+b^2=1$. We used $a=1.00, 0.95, 0.90,$ and $0.87$. The results of performing EATRf and OPESf on these simulations are given in \fig{fig:cavity_lincombo}.
Adding weight to $\theta$ and removing weight from $z$ in the CV is expected to make the CV worse for biasing, and in Fig.~\ref{fig:cavity_lincombo}, we can clearly see the effect of systematically making our biased CV worse, as the slope decreases and deviates from 1, which is the OPESf approximation. 

Performing our full EATRf analysis, we find that $\gamma$ decreases systematically along with the fraction of $\theta$ introduced in Fig.~\ref{fig:cavity_lincombo}b. At the same time, this leads to a decrease in the OPESf rate estimate, i.e. a severe overestimate of the binding lifetime. 
In contrast, we were pleased to see that the rate prediction from EATRf is roughly the same even as the CV gets substantially worse.

\section{Conclusions}
\label{sec:conclusions}

In this study, we developed a variant of the EATR method that allows for a more general way of obtaining accurate rate constants when applying biasing to CVs of non-ideal quality, including methods that apply a quasi-static bias where our previous formulation failed.
The primary innovation is to run many sets of biased simulations, each with a different amount of bias.
These multiple sets of biased data allow us to visualize whether the bias coordinate is efficient. Moreover, we can use the relationship between the amount of bias, the observed rate, and the unbiased rate to estimate the unbiased rate for each set of simulations as a function of the $\gamma$ parameter. The value of $\gamma$ that minimizes the variance in these estimates is taken as the $\gamma$ estimate, and the average of the corresponding rate estimates at that $\gamma$ is taken as the unbiased rate estimate for the method. We found that we can use multiple sets of simulations to measure the sensitivity of the observed rate on the bias, which determines the $\gamma$ parameter for EATR.

For the cases we have evaluated, EATR-flooding produces consistent and accurate rate constant estimates when varying the choice of biased CV or CVs. 
While this requires performing multiple sets of simulations, reasonable rates can be obtained with relatively few simulations per set and a small number of sets such that there is only a relatively small increase in cost for a potentially substantial gain in accuracy.
It was also shown that when enough bias has been added to reach the overbiasing regime, the rate becomes less sensitive to the further addition of bias, allowing us to diagnose when the system is being truly over-biased in a manner that may prevent extracting a true rate.

By extending the EATR method to work with OPES simulations, we have not only improved the capability for rapid biasing, but we have also opened up the possibility to gain control over the biasing with OPES, using its inbuilt limitation in maximum bias, $\Delta E$, and optionally the use of an excluded region for applying bias.
This will allow us to tackle kinetics computations for challenging biomolecular problems, such as in our previous work evaluating the force-dependence of the unbinding of the protein vinculin from actin, where OPESf seemed to provide more stable estimates of kinetics than iMetaD \cite{pena2025direct}.
We would now like to revisit those calculations to determine whether additional OPES simulations varying $\Delta E$ will substantially change our estimates of vinculin binding lifetimes, and whether this approach can allow us to find a better set of CVs for efficient lifetime computations.

As described in the theory section, we are not limited to combining sets of OPES simulations, in principle this should apply to other biasing methods as well, including time-dependent approaches like MetaD. 
While our preliminary results in Fig.~\ref{fig:eatrf_metad_bias} show that EATRf can be applied directly to iMetaD data using, as an example,  different hill addition rates, further work is required to determine whether optimal choices for iMetaD parameters are required.
Moreover, we would like to investigate whether we can combine our two approaches to take advantage of both the increase of bias between sets of simulations and the increase of bias over the course of each simulation to determine the sensitivity of the rate to bias, and thus obtain a more accurate estimate of $\gamma$.

\section{Computational Methods}
\label{sec:methods}
MD simulations were performed in LAMMPS \cite{thompson2022lammps} and GROMACS \cite{abraham2015gromacs} using standard protocols described in the methods. Enhanced sampling simulations using both MetaD and OPES were performed with the PLUMED open source sampling library \cite{tribello2014plumed}. We have previously contributed an example of computing rates of such systems under applied mechanical force in the PLUMED-tutorials resource \cite{tribello2025plumed}.

\subsection{MD Simulation details}

The molecular dynamics simulations of the G\=o-like model of protein G were performed using the 23 Jun 2022 version of the LAMMPS\cite{thompson2022lammps} software package patched with PLUMED 2.8.3.\cite{tribello2014plumed} The OPES simulation data where the $Q$ and $R_\mathrm{ee}$ CVs were used for biasing were the same data as used in the Supporting Information of Ref.~~\onlinecite{mazzaferro2024good}. The data where the $R_\mathrm{g}$ and $R_\mathrm{g}/R_\mathrm{ee}$ CVs were used for biasing were collected using similar scripts. The simulations were run with a 10\ fs time step, using the Nos\'e-Hoover chain thermostat at 312 K with a damping parameter of 1\ ps and a chain length of 3. Each simulation was stopped once the protein unfolded, which we defined to occur once the $Q$ CV dropped below 0.35.

The cavity-ligand model was adapted from the model used in Ref.~~\onlinecite{pena2022assessing}, which was the same as in Ref.~~\onlinecite{ahalawat2020solvent}. The model is composed of a cube of hexagonally close-packed carbon-like atoms with an ellipsoidal cavity and a buckyball (C60) ligand, solvated with TIP4P water molecules. The cavity and ligand atoms have attractive non-bonded interactions. The model used in this project was adjusted to weaken these interactions, as described in the SI.

The MD simulations for the cavity-ligand model were performed using GROMACS 2023.2 \cite{abraham2015gromacs} patched with PLUMED 2.9.\cite{tribello2014plumed}. The simulations were run in a periodic box with a 2\ fs time step at 300 K, using the V-Rescale temperature coupling.  Each simulation ended once the ligand unbound from the cavity or 500\ ns had been simulated. Ligand unbinding was defined to occur once the $z$-distance increased past 15 \AA.

\subsection{EATR-flooding}

The rate estimations were performed on collections of OPES simulation sets, where each simulation set was run with a different value for $\Delta E$, the \texttt{BARRIER} parameter in PLUMED. The protein G simulations were run with $\Delta E$ ranging from 5 to 13 kJ/mol with 2 kJ/mol increments. The excluded regions for the bias in the protein G simulations for each CV were $Q<0.65$, $R_\mathrm{g}>1.15~\mathrm{nm}$, and $R_\mathrm{ee}>2.9~\mathrm{nm}$. For the 2D $R_\mathrm{g}$, $R_\mathrm{ee}$ CV, bias was excluded from the union of the excluded regions for the separate CVs. The cavity-ligand model simulations shown in the main text were run with $\Delta E$ ranging from 3 to 6 kcal/mol in 1 kcal/mol increments, plus 3.5 and 4.5 kcal/mol. Those in the SI for the original model from Ref.~\citenum{tiwary2015role} used $\Delta E$ from 13 to 16 kcal/mol in 1 kcal/mol increments, plus 13.5 and 14.5 kcal/mol. No excluded region was used for the cavity models.

While analyzing the protein G simulation sets, the biased rate constants were determined by fitting the final transition times to a Poisson distribution. The same was done for the original cavity-ligand model. However, for the new cavity-ligand model, we instead used maximum likelihood estimation to determine the biased rate constants because some simulations were stopped prematurely, which would influence the observed rate constant. 

The ensemble average $\langle e^{\beta\gamma V}\rangle$ for a given value for $\Delta E$ was estimated by first averaging the quantity $e^{\beta\gamma V(t)}$ across all running simulations in the set at time $t$, then averaging the result over time. 
This ordering was done to be consistent with our earlier EATR approach, but we also tested doing the calculation  in the other order with minor quantitative differences. 
We applied \eq{eq:opes_flooding_k0} to each simulation set to get a set of $\ln k_0$ estimates as a function of $\gamma$, then minimized the variance in the $\ln k_0$ estimates using the bounded version of Brent's algorithm implemented in the SciPy Python package.\cite{scipy} The minimizing value of $\gamma$ and the mean of the corresponding $\ln k_0$ values are reported as the $\gamma$ and $\ln k_0$ estimates in this method.

\subsection{Error Analysis}
\label{sec:error}

All error values, unless otherwise stated, were obtained by bootstrapping \cite{boostrap}. In this method, we calculate the error in some statistic from a number of datasets constructed by randomly sampling the original dataset with replacement. We resampled 1000 times and calculated the relevant quantity for each resampling (usually the log-rate or $\gamma$), then took the standard deviation across the resamplings as the error in that quantity. The error in the unbiased unfolding rate of the protein G model is a 95\% confidence interval obtained using the bootstrapping method implemented in SciPy.\cite{scipy}

For the estimation of the unbiased rate for the weakly bound cavity model, we used Bayesian analysis to estimate a 95\% credibility interval from our set of simulations where we observed 5 transitions within 360 microseconds. The posterior distribution, $\pi(k_0|N,T_\mathrm{total})$, given a number of transitions $N$ and a total simulation time $T_\mathrm{total}$ was determined following Ref.~~~\onlinecite{ibrahim2001bayesian} using a noninformative gamma prior distribution $\pi(k_0)\propto 1/k_0$, leading to
\begin{equation}
    \pi(k_0|N,T_\mathrm{total})\propto k_0^{N-1}\exp\left(-k_0T_\mathrm{total}\right)\,.
    \label{eq:gamma_prior}
\end{equation}
This was then converted to the posterior distribution for $\ln k_0$. The highest-density credible interval is obtained from this posterior distribution by finding two $\ln k_0$-values $a$ and $b$ where $P(a<\ln k_0<b|N,T_\mathrm{total})=95\%$ and $\pi(\ln k_0=a|N,T_\mathrm{total})=\pi(\ln k_0=b|N,T_\mathrm{total})$.

\section*{Data availability}
Data and analysis scripts are available from the GitHub repository for this paper, \url{https://github.com/hocky-research-group/EATRf-paper-2026}. The general EATR analysis framework software is available from \url{https://github.com/hocky-research-group/EATR-rate-analysis}.

\section*{Acknowledgments}

NM and GMH were supported by the National Institutes of Health under award number R35GM138312. NM was also supported by a fellowship from the Simons Center for Computational Physical Chemistry (SCCPC, Simons Foundation grant MPS-T-MPS-00839534, MET) and a Margaret Strauss Kramer Fellowship from the Department of Chemistry at NYU. This work was supported in part through the NYU IT High Performance Computing resources, services, and staff expertise, and simulations were partially executed on resources supported by the SCCPC.
The Flatiron Institute is a division of the Simons Foundation. GMH and NM thank the Center for Computational Mathematics at the Flatiron Institute for their hospitality while a portion of this research was carried out.

\bibliography{eatr-flooding}

\clearpage
\onecolumn

{\par\centering\Huge Supporting Information \centering\par}

\renewcommand{\vec}[1]{{\bf{#1}}}

\renewcommand{\thefigure}{S\arabic{figure}}
\setcounter{figure}{0}
\renewcommand{\thesection}{S\arabic{section}}
\setcounter{section}{0}

\section{Further results for protein G}
\begin{figure*}
    \centering
    \includegraphics[width=\textwidth]{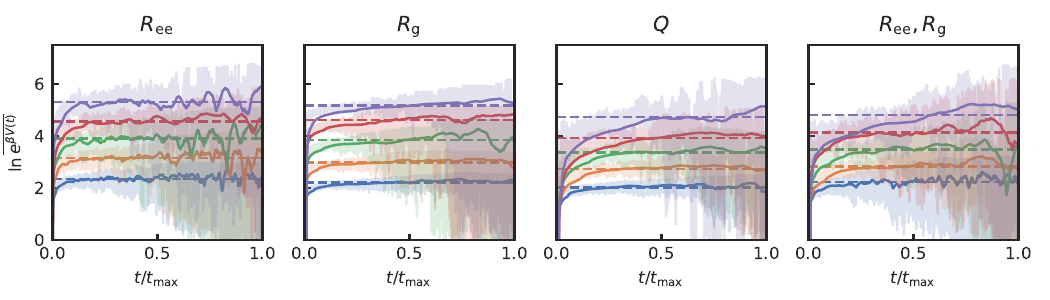}
    \caption{The log of the value of $e^{\beta V(t)}$ averaged over all simulations in each set (low opacity). The solid lines represent the log of the time average of $\overline{e^{\beta V(t)}}$ over a window of 25000 steps, and the horizontal dashed lines represent the time average from 0 up to the maximum transition time, which approximates the log of the ensemble average $\ln\left<e^{\beta V}\right>$ in the folded state.}
    \label{fig:opes_biasmeasure}
\end{figure*}

\begin{figure*}
    \centering
    \includegraphics[width=\textwidth]{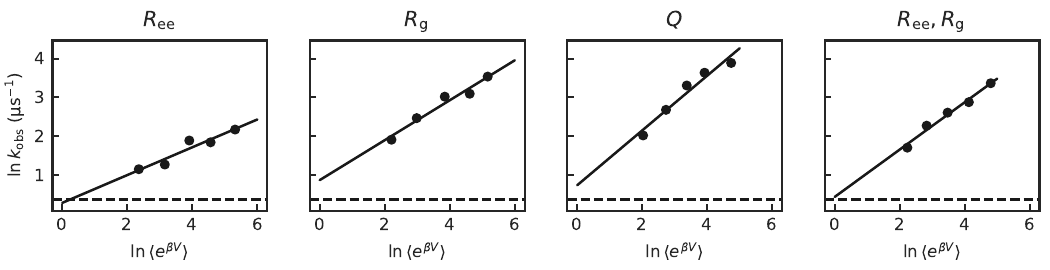}
    \caption{The logarithm of the observed rate plotted against $\ln\left<e^{\beta V}\right>$ in the folded state of the protein G model, estimated as the log of the time average of $\overline{e^{\beta V(t)}}$. The horizontal dashed line represents the value of $\ln k_0$. Increasing $\ln\left<e^{\beta V}\right>$ for poor CVs has less of an effect on $\ln k_\mathrm{obs}$ than for good CVs. The EATR-slope approach takes the y-intercept of the best fit line to be the estimate for $\ln k_0$ and the slope to be the estimate for $\gamma$.}
    \label{fig:opes_slope_protg}
\end{figure*}

\subsection{Static biasing}
\label{sec:static_bias}

\begin{figure*}
    \centering
    \includegraphics[width=\textwidth]{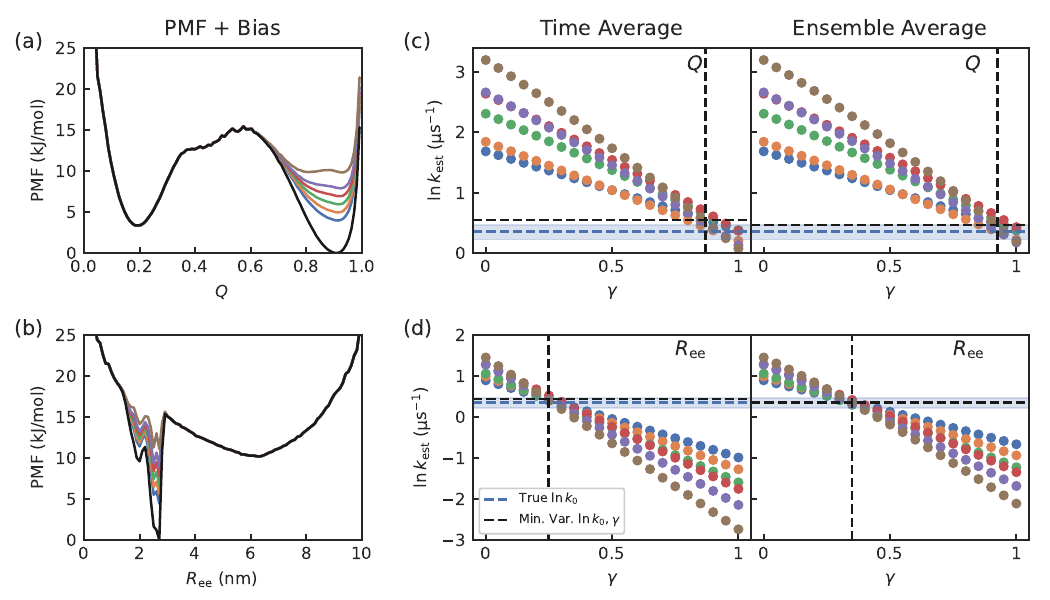}
    \caption{(a), (b) The PMFs of the protein G model for $Q$ and $R_\mathrm{ee}$. The unbiased PMF is shown in black while the biased PMF for each bias level is colored. (c), (d) The values of $\ln k_0$ predicted by \eq{eq:opes_flooding_k0} for each set of simulations for various values of $\gamma$, by approximating the expected value $\langle e^{\beta\gamma V}\rangle$ using the time average of the simulation average (left) and the directly estimated ensemble average (right). }
    \label{fig:eatrf_static_bias}
\end{figure*}

Because the EATR-flooding theory requires the bias to be effectively static, we decided to also run the analysis on a collection of simulation sets with exactly static biasing potentials. We ran six sets of simulations each for the $Q$ and $R_\mathrm{ee}$ CVs of the protein G system, each with nominal bias levels of 4, 5, 6, 7, 8, and 10 kJ/mol.

For the $Q$ CV, we applied a bias potential of the form
\begin{equation*}
    V(Q)=H\exp\left( -(Q-\mu)^2/a^2 \right)\,,
\end{equation*}
where $H$ is the nominal bias level. We used $\mu=0.9$ and $a=0.13$.

Using the same EATR-flooding procedure as in the main text, we obtain $\ln k_0=0.53\pm0.16$ and $\gamma=0.88\pm0.07$. The biased PMFs and the values of $\ln k_0$ obtained from \eq{eq:opes_flooding_k0} are given in \fig{fig:eatrf_static_bias}a,c.

Because we have the exact expression for the bias, as well as a good estimate for the potential of mean force (PMF) in the folded state, we decided to also directly calculate the $\ln \langle e^{\beta\gamma V}\rangle$ term in \eq{eq:opes_flooding_k0} using
\begin{equation*}
    \langle A\rangle\approx\frac{\sum_i A(\xi_i)e^{-\beta(F(\xi_i)+V(\xi_i))}}{\sum_i e^{-\beta(F(\xi_i)+V(\xi_i))}}\,,
\end{equation*}
where $\xi_i$ are evenly spaced values for the CV $\xi$ in the reactant state and $F(\xi_i)$ is the PMF of $\xi$ at $\xi_i$. Using this instead yields $\ln k_0=0.48\pm0.14$ and $\gamma=0.92\pm0.06$.

For the $R_\mathrm{ee}$ CV, we applied a bias potential of the form
\begin{equation*}
    V(R_\mathrm{ee})=H\left(0.9\exp\left( -(R_\mathrm{ee}-\mu_1)^2/a_1^2 \right)+\exp\left( -(R_\mathrm{ee}-\mu_2)^2/a_2^2 \right)+0.5\exp\left( -(R_\mathrm{ee}-\mu_3)^2/a_3^2 \right)\right)\,,
\end{equation*}
where we used $\mu_1=2.45~\mathrm{nm}$, $\mu_2=2.68~\mathrm{nm}$, $\mu_3=1.9~\mathrm{nm}$, $a_1=0.2~\mathrm{nm}$, $a_2=0.12~\mathrm{nm}$, and $a_3=0.3~\mathrm{nm}$.

Estimating $\ln \langle e^{\beta\gamma V}\rangle$ as in the main text results in $\ln k_0=0.40\pm0.18$ and $\gamma=0.31\pm0.07$. Estimating it as above instead results in $\ln k_0=0.34\pm0.18$ and $\gamma=0.35\pm0.07$. The biased PMFs and the values of $\ln k_0$ obtained from \eq{eq:opes_flooding_k0} are given in \fig{fig:eatrf_static_bias}b,d.

\subsection{Metadynamics biasing}
\label{sec:metad_bias}

\begin{figure*}
    \centering
    \includegraphics[width=\textwidth]{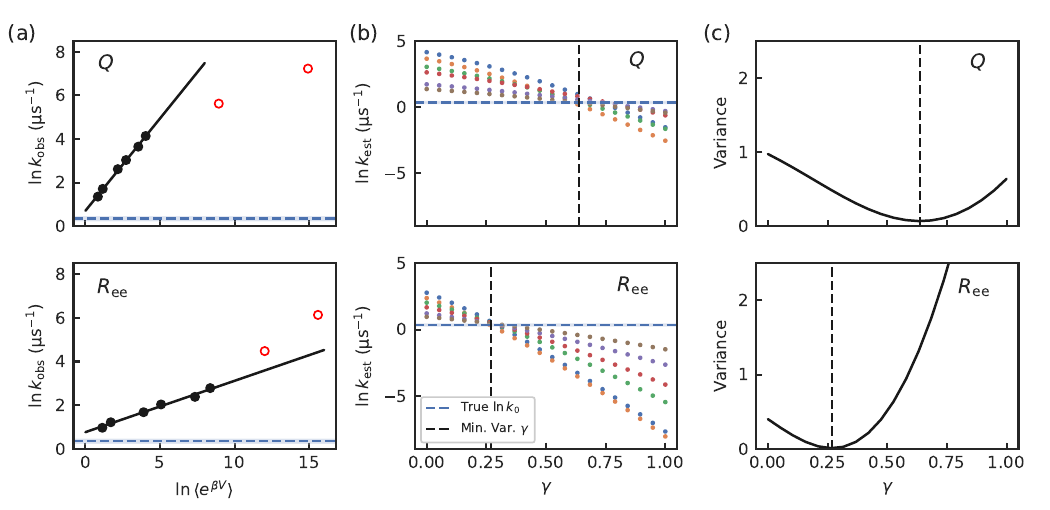}
    \caption{(a) The observed rate constant $\ln k_\mathrm{obs}$ plotted against the $\ln \left< e^{\beta V} \right>$ in the folded state of the protein G model, estimated as the average acceleration factor $\ln \overline{\alpha_i}$. The red open points are the sets of simulations which deviate from linearity, and are discarded. (b) The values of $\ln k_0$ predicted by Eq.~\ref{eq:opes_flooding_k0} for each set of simulations for various values of $\gamma$ (c) The variance in the predicted value of $\ln k_0$ across the sets of simulations. The horizontal dashed lines represent the true value of $\ln k_0$, with the 95\% confidence interval represented by the blue shaded region. The vertical dashed lines represent the minimum variance value of $\gamma$. }
    \label{fig:eatrf_metad_bias}
\end{figure*}

We also decided to run the analysis on a set of MetaD simulations to see if this procedure also works for non-static biasing potentials. We applied the EATR-flooding approach to the MetaD simulations of the protein G model which were run with different bias deposition paces and were analyzed in Ref.\citenum{mazzaferro2024good}. The simulations were run using well-tempered MetaD using 10.0 for the BIASFACTOR parameter, with a hill height of 0.4 kJ/mol and $\sigma$ of 0.06 nm for $R_\mathrm{ee}$, and a hill height of 0.6 kJ/mol and $\sigma$ of 0.02 for $Q$. Each set of simulations was run using a different PACE parameter, from 1 ps to 10 ns per hill deposition.

First, the linear fit of $\ln k_\mathrm{obs}$ as a function of $\ln \left< e^{\beta V} \right>$ was assessed as shown in \fig{fig:eatrf_metad_bias}a. The value of $\ln k_\mathrm{obs}$ was obtained by taking the reciprocal of the mean transition time, as the transition time distribution was expected to be non-Poissonian and therefore fitting the CDF would not be appropriate. The value of $\ln \left< e^{\beta V} \right>$ was estimated by taking the final value of the acceleration factor for each simulation

\begin{equation*}
    \alpha_i=\frac{1}{t_i}\int_0^{t_i}e^{\beta V(\xi (t'),t')}~dt'
\end{equation*}
and averaging across all simulations in the set. The fastest paces corresponding to deposition times of 1 and 10 ps were observed to deviate from the expected linear behavior, and were thus discarded.

The EATR-flooding protocol, where $\ln \left< e^{\beta\gamma V} \right>$ was estimated by taking the simulation average before the time average, was then applied to the remaining data, as shown in \fig{fig:eatrf_metad_bias}b,c. For $Q$, the least-variance value for $\gamma$ was $0.67\pm0.06$ and the corresponding value for $\ln k_0^*$ was $0.47\pm0.17$ with $k_0^*$ in ($\mathrm{\mu s}^{-1}$). For $R_\mathrm{ee}$, $\gamma^*$ was $0.28\pm0.01$ and the corresponding value for $\ln k_0^*$ was $0.43\pm0.06$.

\section{Further results for the cavity-ligand model}

\label{sec:si_cavity}
The receptor is a semi-hollow cube that has a cavity which is half an ellipsoid of radius ~11 \AA. Moreover, the cube consists of two types of hydrophobic atoms; the cavity atoms (CP) and the wall atoms (CW) (the atoms in the cavity have a higher attraction to the ligand atoms (CF) than do the wall atoms), and the whole complex model is solvated with TIP4P water.  
Parameters for non-bonded interactions between receptor and ligand atoms were determined using GROMACS' combination rule 3, where $\sigma_{ij} = \sqrt{\sigma_i \sigma_j}$ and $\epsilon_{ij} = \sqrt{\epsilon_i \epsilon_j}$.  The $\epsilon$ values were reduced while keeping the ratio between $\epsilon_{\mathrm{CP-CF}}$ and $\epsilon_{\mathrm{CW-CF}}$ fixed.
\begin{table}
\begin{center}
\caption{\label{tab:lj} Lennard-Jones parameters for the receptor and ligand atoms. CW, CP, and CF refer to the wall, cavity, and fullerene atoms respectively }
\begin{tabular}{lcccr}
\hline
atom-type&$\sigma ({\rm nm})$&$\epsilon (\frac{{\rm kJ}}{{\rm mol}})$\\
\hline
CF & 0.3500 & 0.276144  \\
CW & 0.4152 & 0.002400  \\
CP & 0.4152 & 0.008000  \\
\end{tabular}
\end{center}
\end{table}

\begin{table*}
    
\begin{center}
\caption{\label{tab:ljnb} Lennard-Jones parameters for non-bonded interactions between receptor and ligand atoms. Parameter $\epsilon_1$ was used in the original model, while $\epsilon_3$ was used for the fast unbinding model. The non-bonded interactions of the CP-CP, CW-CP, and CW-CW pairs were excluded from the total energy by setting their LJ parameters to 0 and the entire lattice was position restrained with a $1000\, \frac{kJ}{mol\,nm^2 }$ force constant.}
\begin{tabular}{lcccccr}
\hline
i&j&$\sigma ({\rm nm})$&$\epsilon_1 (\frac{{\rm kJ}}{{\rm mol}})$ &$\epsilon_2 (\frac{{\rm kJ}}{{\rm mol}})$&$\epsilon_3 (\frac{{\rm kJ}}{{\rm mol}})$&$\epsilon_4 (\frac{{\rm kJ}}{{\rm mol}})$\\
\hline
CF & CP & 0.3812 & 0.04700 & 0.03600 & 0.02500 & 0.01850 \\
CF & CW & 0.3812 & 0.02574 & 0.01972 & 0.01369 & 0.01015 \\
CW & CW & 0.4152 & 0.00000 & 0.00000 & 0.00000 & 0.00000 \\
CP & CP & 0.4152 & 0.00000 & 0.00000 & 0.00000 & 0.00000 \\
CP & CW & 0.4152 & 0.00000 & 0.00000 & 0.00000 & 0.00000 \\
\end{tabular}
\end{center}
\end{table*}

\begin{figure}[h]
\includegraphics[]{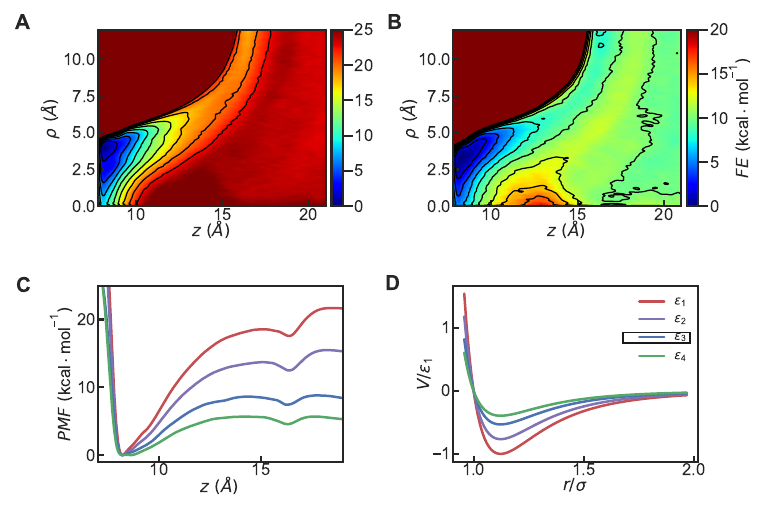}
\caption{\label{fig:wcavitys} FES of original (A) and fast unbinding model, $\epsilon_{3}$ (B), as function of $\rho$ and $z$. (C) Projection of FES onto transverse distance, $z$, for all models. (D) LJ potential for interactions between cavity and ligand atoms (CP-CF) in original and modified models, potential was scaled by $\epsilon_1$ (see Table \ref{tab:ljnb}) and distance was scaled by $\sigma$.}
\end{figure}

In Fig. \ref{fig:wcavitys}C,D, Topology 1 (top. 1) describes the original model and has the highest barrier to unbinding on the $z$ CV. Topology 2 (top. 2) has a lower barrier but remains high and kinetics were still too slow. Topology 4 has the lowest barrier, but the timescale of the unbinding kinetics was reduced to less than a microsecond. Thus, the second-weakest model (top. 3) was chosen for benchmarking.

Using the WTMetaD approach, FES calculations were performed for all models, including the original, biasing both the transverse distance, $z$, and radial distance, $\rho$. The biasing parameters were: HEIGHT=0.478, SIGMA=0.3,0.1, BIASFACTOR=10, and PACE=300. for the modified models. For the original model, only one parameter was different, BIASFACTOR=12, which was higher to obtain comparable sampling within 100 ns. Restraining potentials (PLUMED UPPER\_WALLS) were applied for the transverse and radial distance CVs at 21 \AA\ and 12 \AA, respectively to limit sampling within the periodic box.

\begin{figure*}[h]
    \centering
    \includegraphics[width=\textwidth]{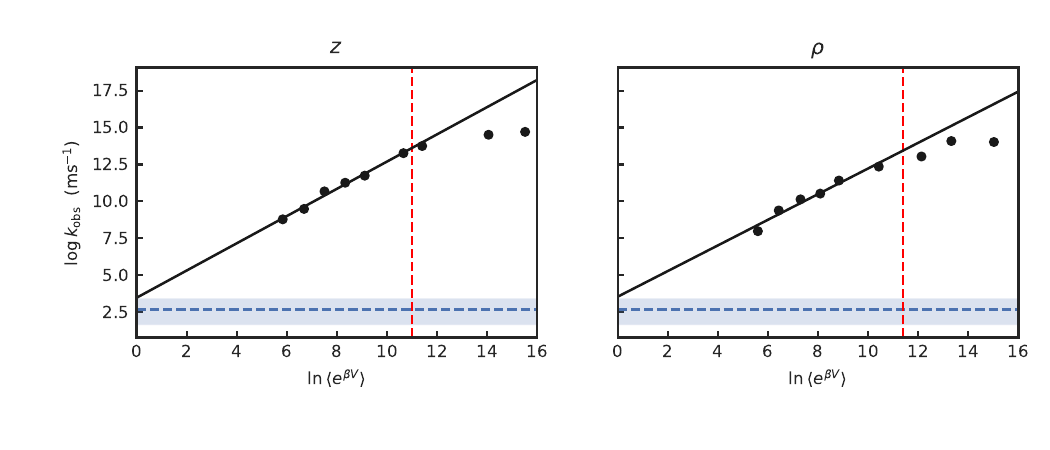}
    \caption{The logarithm of the observed rate plotted against $\ln\left<e^{\beta V}\right>$ in the bound state of the cavity-ligand model, estimated as the log of the time average of $\overline{e^{\beta V(t)}}$. The horizontal dashed line represents the value of $\ln k_0$. Only points to the left of the red vertical dashed line were used in the EATRf analysis in the main text, as the rate becomes less sensitive to bias past that point, demonstrated by the deviation from linear behavior.}
    \label{fig:opes_slope_cavity}
\end{figure*}

\section{Original cavity-ligand model results}
\label{sec:si_cavity_results}

\begin{figure*}
    \centering
    \includegraphics[width=0.75\textwidth]{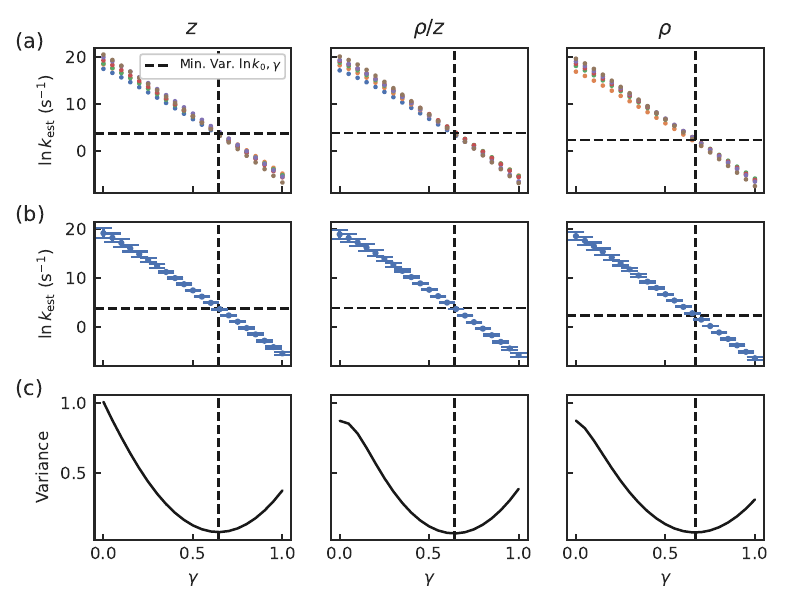}
    \caption{The results of EATRf on the original cavity-ligand model. (a) The values of $\ln k_0$ predicted by \eq{eq:opes_flooding_k0} for each set of OPES simulations with each BARRIER parameter for various values of $\gamma$. (b) The value of $\ln k_0$ averaged over the sets of simulations for each potential value of $\gamma$. The variance of $\ln k_0$ across the sets is represented by error bars. (c) The variance in the predicted value of $\ln k_0$ across the sets of simulations. The dashed lines represent the minimum variance value of $\gamma$, which is the value that causes all sets to give the same value of $\ln k_\mathrm{est}$, and the corresponding $\ln k_0^*$.}
    \label{fig:opes_gamma_cavity}
\end{figure*}

For the original cavity-ligand model, we performed simulations with a set of $\Delta E$ values where we biased $z$ alone, both $z$ and $\rho$, and $\rho$ alone. For all collections of simulation sets, we ran 6 sets of simulations with $\Delta E$ varying from 13 to 16 kcal/mol, except we did not run the 13 kcal/mol set for $\rho$ alone. Because of a few outliers in the data set, the observed unbinding rates in the simulations could not be estimated using the maximum likelihood estimator. $\ln k_\mathrm{obs}$ was estimated by fitting the CDF instead, where in the case where a simulation ended early without transitioning, the CDF is fit only up to the time where that simulation ended. The results from EATRf are given in \fig{fig:opes_gamma_cavity}. The minimum-variance residence time we obtained from EATRf is $\tau_0=0.023~\mathrm{s}$ ($\gamma=0.64$) when biasing $z$ alone, $\tau_0=0.021~\mathrm{s}$ ($\gamma=0.64$) when biasing $z$ and $\rho$ together, and $\tau_0=0.097~\mathrm{s}$ ($\gamma=0.67$) when biasing $\rho$ alone. This is between the residence time estimates from Markov state modeling ($\tau_0=0.00055~\mathrm{s}$)\cite{ahalawat2020solvent} and previous infrequent metadynamics studies ($\tau_0=3863~\mathrm{s}$, $\tau_0=2000~\mathrm{s}$).\cite{tiwary2015role,mondal2013hydrophobic}


\end{document}